\documentclass[aps,prd,twocolumn,superscriptaddress,showkeys,nofootinbib]{revtex4-1}

\usepackage[utf8]{inputenc}
\usepackage{amsmath}
\usepackage{graphicx}
\usepackage{xcolor}

\def\fs#1{\setbox0=\hbox{$#1$}#1\hskip-\wd0\dimen0=5pt\advance\dimen0
by-\ht0\advance\dimen0 by\dp0\lower0.5\dimen0\hbox to\wd0{\hss\sl/\/\hss}}

\begin{document}

\title{Nonrelativistic model of tetraquarks and predictions for their masses\\ from fits to charmed and bottom meson data}

\author{Per Lundhammar}
\email[]{lundhamm@kth.se}
\affiliation{Department of Physics, School of Engineering Sciences, KTH Royal Institute of Technology, AlbaNova University Center, Roslagstullsbacken 21, SE--106~91 Stockholm, Sweden}

\author{Tommy Ohlsson}
\email[]{tohlsson@kth.se}
\affiliation{Department of Physics, School of Engineering Sciences, KTH Royal Institute of Technology, AlbaNova University Center, Roslagstullsbacken 21, SE--106~91 Stockholm, Sweden}
\affiliation{The Oskar Klein Centre for Cosmoparticle Physics, AlbaNova University Center, Roslagstullsbacken 21, SE--106 91 Stockholm, Sweden}

\begin{abstract}
We investigate a nonrelativistic model of tetraquarks, which are assumed to be compact and to consist of diquark-antidiquark pairs. We fit, for the first time, basically all currently known values for the measured masses of 45 mesons, including both charmed and bottom mesons, to the model and predict masses of tetraquarks as well as diquarks. In particular, we find masses of four axial-vector diquarks, i.e., $qc$, $cc$, $qb$, and $bb$, where $q = u,d$, and 24 ground-state tetraquarks, including both heavy-light tetraquarks ($qc\overline{qc}$ and $qb\overline{qb}$) and heavy tetraquarks ($cc\overline{cc}$ and $bb\overline{bb}$). In general, our results for the masses of $qb\overline{qb}$, $cc\overline{cc}$, and $bb\overline{bb}$ are largely comparable with other reported results, whereas our results for the masses of $qc\overline{qc}$ are slightly larger than what has been found earlier. Finally, we identify some of the obtained predictions for masses of tetraquarks with masses of experimental tetraquark candidates, and especially, we find that $\psi(4660)$, $Z_b(10610)$, and $Z_b(10650)$ could be described by the model.
\end{abstract}

\maketitle

\section{Introduction\label{sec:intro}}

The concept of hadrons was introduced in 1962 by Okun \cite{Okun:1962kca} and developed into the quark model in 1964 independently by Gell-Mann \cite{GellMann:1964nj} and Zweig \cite{Zweig:1981pd, Zweig:1964jf}, describing ordinary mesons ($q\bar{q}$) and baryons ($qqq$) in terms of quarks $q$ and antiquarks $\bar{q}$. In addition to the quark model, the possible existence of exotic hadrons, such as tetraquarks ($q\bar{q}q\bar{q}$ or $qq\bar{q}\bar{q}$) and pentaquarks ($q\bar{q}qqq$), consisting of four or more quarks was proposed in Gell-Mann's seminal work \cite{GellMann:1964nj}, but it was not until the beginning of the 21st century when the first claimed observations of exotic hadrons were made \cite{Jaffe:2004ph}. Today, a large amount of data, obtained at both electron-positron and hadron colliders, has provided evidence for the possible existence of such exotic hadrons. Concerning tetraquarks, the first discovery was made in 2003 by the Belle Collaboration that observed a resonance peak at ($3872.0 \pm 0.6$)~MeV \cite{Choi:2003ue}, which was named $X(3872)$ (and now sometimes referred to as $\chi_{c1}(3872)$ \cite{Zyla:2020}), and then confirmed by several other experiments (see, e.g., the reviews~\cite{Swanson:2006st, Godfrey:2008nc} and references therein). Many proposed exotic hadrons only appear in one decay mode, although $X(3872)$ can be observed in several other decay modes as was discovered by the BaBar \cite{Aubert:2004fc}, CDF \cite{Acosta:2003zx}, and D$\emptyset$ \cite{Abazov:2004kp} Collaborations. Later, the ATLAS, CMS, and LHCb Collaborations were able to contribute with a massive amount of data on the electrically charge neutral $X(3872)$ and its current mass is determined to be ($3871.69 \pm 0.17$)~MeV \cite{Zyla:2020, Cowan:2018fki}. It is the most studied exotic hadron, but its nature is still fairly unknown. It has similar properties to the charmonium state $c\bar{c}$ and was first believed to be an undiscovered excited state of $c\bar{c}$, but a closer investigation of the decay modes $X(3872) \to J/\psi \, \pi^+ \, \pi^-$ and $X(3872) \to J/\psi \, \omega$ shows violation of isospin \cite{Abe:2005ix, delAmoSanchez:2010jr}, which is unusual for $c\bar{c}$. If $X(3872)$ is an exotic hadron, then a common description is that it contains two quarks and two antiquarks forming $uc\overline{uc} $. However, it is still an open problem how it is bound together. Since the discovery of $X(3872)$, many new exotic hadron candidates have been claimed to be observed with final states of a pair of heavy quarks and a pair of light antiquarks, which are labeled as $X$, $Y$, and $Z$ states by experimental collaborations and collectively referred to as $XYZ$ states \cite{Brambilla:2019esw}. Examples of candidates for $XYZ$ states are $Z_c(3900)$ \cite{Ablikim:2013mio,Liu:2013dau}, $Z_c(4025)$ \cite{Ablikim:2013emm, Ablikim:2013wzq} (now known as $X(4020)^\pm$ \cite{Zyla:2020}), $Z_b(10610)$ \cite{Oswald:2013tna}, and $Z_b(10650)$ \cite{Oswald:2013tna}. The dynamics of the $XYZ$ systems involves both short and long distance behaviors of QCD, which make theoretical predictions difficult. Hence, many competing phenomenological models currently exist for such states, including lattice QCD, compact tetraquark states, molecular states, QCD sum rules, coupled-channel effects, dynamically generated resonances, and nonrelativistic effective field theories (see Ref.~\cite{Ghalenovi:2020zen} and references therein). Many models view the exact nature of the inner structure of tetraquarks to be compact and to consist of so-called diquark-antidiquark pairs \cite{Ali:2019roi}. A diquark is a bound quark-quark pair, whereas an antidiquark is a bound antiquark-antiquark pair. These pairs are not by themselves colorless, but are proposed in the context of tetraquarks to form colorless combinations. Exotic hadrons of such pairs are thus not ruled out by QCD, but cannot be accommodated within the naive quark model. Modeling of tetraquarks containing only heavy quarks is therefore of special interest and easier to study theoretically, since several assumptions can be justified. Recently, the LHCb Collaboration reported the observation of a doubly charmed and doubly charged baryon $\Xi_{cc}^{++}$ \cite{Aaij:2017ueg}, which has lead to further attention on heavy-quark systems as the description of exotic hadrons. Many tetraquark candidates are not possible to describe within quark models, since they have electric charge, and therefore cannot be charmonium or bottomonium, but are potential candidates for hidden-charm or hidden-bottom tetraquarks, molecular systems \cite{Esposito:2016noz,Guo:2017jvc} of charmed or bottom mesons or hadroquarkonia (see Ref.~\cite{Bedolla:2019zwg} and references therein).

In this work, we will study a nonrelativistic model describing tetraquarks as composed of diquarks and antidiquarks, which interact much like ordinary quarkonia. Performing numerical fits to masses of mesons, masses of tetraquarks and the underlying diquarks will be predicted. For the first time, we will use both charmed and bottom mesons in the same fit and data of in total 45 charmed and bottom mesons (e.g., charmonium, bottomonium, $D$ mesons, and $B$ mesons) will be considered. We will predict masses of 24 tetraquark states, which is more than what has previously been performed in the literature.

This work is organized as follows. In Sec.~\ref{sec:model}, we present the nonrelativistic diquark-antidiquark model describing tetraquarks that can predict their masses and describe the numerical fitting procedure of meson data. Then, in Sec.~\ref{sec:results}, we perform numerical fits of this model and state the results of the fits, including predicted masses of diquarks and tetraquarks. We will also present a thorough discussion on the results obtained and comparisons to other works, both theoretical and experimental. Finally, in Sec.~\ref{sec:summary}, we summarize our main results and state our conclusions.

\section{Model and fitting procedure}
\label{sec:model}

In this section, the model of tetraquarks viewed as diquark-antidiquark systems is presented, and the method used to prescribe some tetraquark states quantitative masses is derived. This is preformed by firstly considering a quark-antiquark system and describing the Hamiltonian of that system with an unperturbed one-gluon exchange (OGE) potential and a perturbation term taking the spin of the system into account. This gives rise to a model with four free parameters, which are then fitted to meson data. Secondly, the model is expanded to incorporate composite quark-quark systems, which are called diquarks (the antiquark-antiquark systems are called antidiquarks). Thirdly, with the masses of the diquarks determined, the initial stage of the model describing quark-antiquark systems is then used to describe the diquark-antidiquark systems, which are interpreted as bound states of tetraquarks. 

\subsection{Model procedure}

\begin{figure}[!t]
    \centering
    \includegraphics[width=0.45\textwidth]{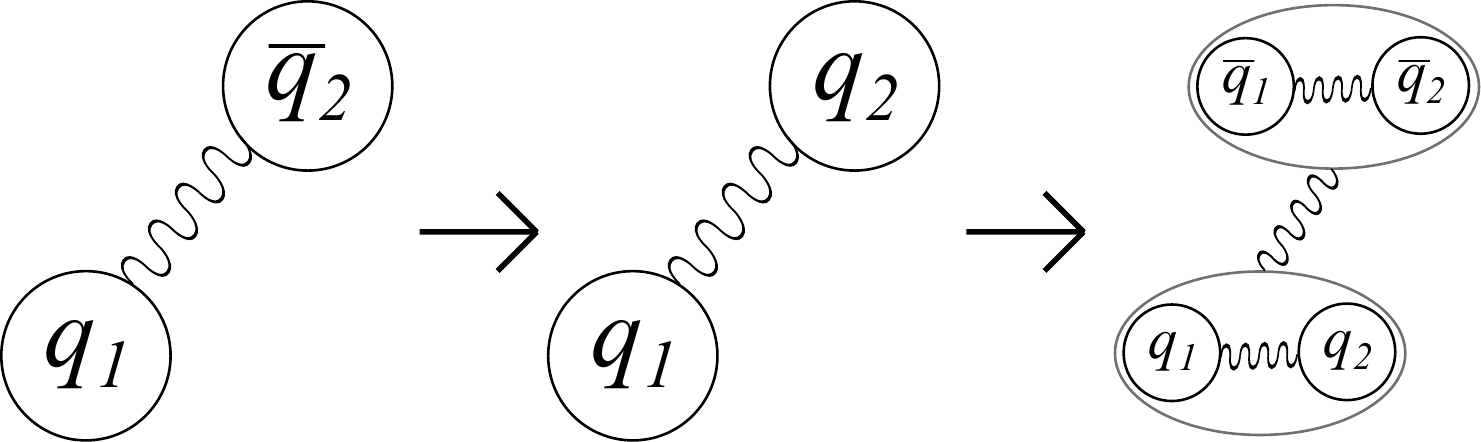}
    \caption{A schematic overview of the modeling procedure. First, considering a model of the quark-antiquark system $q_1\bar{q}_2$. Second, extrapolating the model to also describe the quark-quark (or diquark) system $q_1q_2$. Third, modeling a tetraquark $q_1 q_2 \overline{q_1 q_2}$ in the same way as the quark-antiquark system, but with diquarks and antidiquarks as constituents.}
    \label{fig:TetraquarkSchematics}
\end{figure}

The modeling procedure can be outlined and summarized as follows:
\begin{enumerate}
    \item Fitting a quark-antiquark model to meson data to obtain the parameters of the effective potential.
    \item Using that set of parameters to determine the diquark and antidiquark masses by changing the color constant and the string tension of the potential.
    \item Considering the diquarks and antidiquarks as constituents of the tetraquarks to predict the tetraquark masses, see Fig.~\ref{fig:TetraquarkSchematics} for a schematic overview of the modeling procedure. 
\end{enumerate}
We begin by considering the interaction between a quark and an antiquark. In quark bound state spectroscopy, a commonly used potential describing the unperturbed contribution is the so-called Cornell potential \cite{Eichten:1979ms},
\begin{equation}
    V(r) = \frac{\kappa\alpha_S}{r} + br \,,
    \label{eq:Cornell}
\end{equation}
where $\kappa$ is a color factor and associated with the color structure of the system, $\alpha_S$ the fine-structure constant of QCD, and $b$ the string tension. The first term in Eq.~(\ref{eq:Cornell}), i.e., $V_V(r)\equiv\kappa\alpha_S/r$, is the Coulomb term and associated with the Lorentz vector structure. It arises from the OGE between the quarks. The second term in Eq.~(\ref{eq:Cornell}) is associated with the confinement of the system. A nonrelativistic approach is legitimate under the condition that the kinetic energy is much less than the rest masses of the constituents, which is usually the case considering heavy-quark bound states. We formulate the Schr{\"o}dinger equation in the center-of-mass frame. Using spherical coordinates, one can factorize the angular and radial parts of this Schr{\"o}dinger equation. Now, let $\mu\equiv m_1m_2/(m_1+m_2)$, where $m_1$ and $m_2$ are the constituent masses of quark~1 and quark~2, respectively. In the case that $m \equiv m_1 = m_2$, it holds that $\mu = m/2$. Thus, the time-independent radial Schr{\"o}dinger equation can be written as
\begin{equation}
    \left\{  -\frac{1}{2\mu}\left[\frac{\text{d}^2}{\text{d}r^2}+\frac{2}{r}\frac{\text{d}}{\text{d}r}-\frac{L(L+1)}{r^2}\right]+V(r) \right\} \psi(r) = E\psi(r) \,,
    \label{eq:Schrodinger}
\end{equation}
with the orbital quantum number $L$ and the energy eigenvalue $E$. Substituting $\psi(r) \equiv r^{-1}\varphi(r) $, Eq.~(\ref{eq:Schrodinger}) transforms into 
\begin{equation}
    \left\{  \frac{1}{2\mu}\left[-\frac{\text{d}^2}{\text{d}r^2}+\frac{L(L+1)}{r^2}\right] +V(r) \right\} \varphi(r) = E\varphi(r) \,.
    \label{eq:SchrodingerRed}
\end{equation}
Based on the Breit--Fermi Hamiltonian for OGE, one can include a spin-spin interaction on the form \cite{Lucha:1991vn, Lucha:1995zv, Voloshin:2007dx, Eiglsperger:2007ay, Debastiani:2017msn},
\begin{align}
    V_S(r) &= -\frac{2}{3(2\mu)^2}\nabla^2V_V(r)\mathbf{S}_1\cdot\mathbf{S}_2 \nonumber\\
    &= -\frac{2\pi\kappa\alpha_S}{3\mu^2}\delta^3(r)\mathbf{S}_1\cdot\mathbf{S}_2 \,.
    \label{eq:spin1}
\end{align}
In this model, we incorporate the spin-spin interaction $V_S(r)$ in the unperturbed potential $V(r)$ by replacing the Dirac delta function with a smeared Gaussian function, depending on the parameter $\sigma$, in the following way:
\begin{equation}
    V_S(r) = -\frac{2\pi\kappa\alpha_S}{3\mu^2}\left(\frac{\sigma}{\sqrt{\pi}}\right)^3\exp\left(-\sigma^2 r^2\right)\mathbf{S}_1\cdot\mathbf{S}_2,
    \label{eq:spin2}
\end{equation}
as performed in Ref.~\cite{Godfrey:1985xj}. Now, Eq.~(\ref{eq:SchrodingerRed}) takes the simple form,
\begin{equation}
    \left[  -\frac{\text{d}^2}{\text{d}r^2} +V_{\text{eff}}(r) \right]\varphi(r) = 2\mu E\varphi(r) \,,
    \label{eq:SchrodingerRed2}
\end{equation}
where the effective potential $V_{\text{eff}}(r)$ is given by
\begin{equation}
    V_{\text{eff}}(r)\equiv 2\mu\left[ V(r)+V_S(r) \right] + \frac{L(L+1)}{r^2} \,,
    \label{eq:spin3}
\end{equation}
taking into account the spin-spin interaction. Equation~(\ref{eq:SchrodingerRed2}) can be solved numerically for the energy eigenvalue $E$ and the reduced wave function $\varphi(r)$. The mass $M$ of the bound quark-antiquark system can then be expressed as 
\begin{equation}
    M = m_1+m_2+E \,.
\end{equation}
Note that this model consists of five unknown free parameters, namely the masses $m_1$ and $m_2$ of the two constituents, the fine-structure constant $\alpha_S$ of QCD, the string tension $b$, and the parameter $\sigma$ of the spin-spin interaction.

\subsection{Color structure}
\label{sub:color}

Hadrons are only stable when the colors of their constituent quarks sum up to zero, and thus, every naturally occurring hadron is a color singlet under the group symmetry $\mbox{SU(3)}$. This means that a hadron only occurs if the product color state of the constituent quarks decomposes to an irreducible representation with dimension equal to one.

The difference in color structure between the quark-antiquark and quark-quark systems allows us to extend the model of the quark-antiquark system to also be valid considering a quark-quark system by only changing the color factor $\kappa$ and the string tension $b$. The $\mbox{SU(3)}$ color symmetry of QCD implies that the combination of a quark and an antiquark in the fundamental color representation can be reduced to $|q\bar{q}\rangle :\mathbf{3}\otimes\mathbf{\bar{3}} = \mathbf{1}\oplus\mathbf{8}$,
which gives the resulting color factor for the color singlet as $\kappa=-4/3$ for the quark-antiquark system. When combining two quarks in the fundamental color representation, it reduces to $|qq\rangle :\mathbf{3}\otimes\mathbf{3} = \mathbf{\bar{3}}\oplus\mathbf{6}$, i.e., a color antitriplet $\mathbf{\bar{3}}$ and a color sextet $\mathbf{6}$. Similarly, when combining two antiquarks, it reduces to a triplet $\mathbf{{3}}$ and an antisextet $\mathbf{\bar{6}}$. Furthermore, combining an antitriplet diquark and a triplet antidiquark yields $|[qq]-[\overline{qq}]\rangle :\mathbf{3}\otimes\mathbf{\bar{3}} = \mathbf{1}\oplus\mathbf{8}$, thus forming a color singlet for which the Coulomb part of the potential is attractive. The antitriplet state is attractive and has a corresponding color factor of $\kappa=-2/3$, while the sextet state is repulsive and a color factor of $\kappa=+1/3$. Therefore, we only consider diquarks in the antitriplet state. Thus, the effect of changing from a quark-antiquark system with the color factor $\kappa=-4/3$ to a diquark system with the color factor $\kappa=-2/3$ is equivalent of introducing a factor of $1/2$ in the Coulomb part of the potential for the quark-antiquark system. It is common to view this factor of $1/2$ as a global factor, since it comes from the color structure of the wave function, thus also dividing the string tension $b$ by a factor of 2. For further details, see Ref.~\cite{Debastiani:2017msn}. We apply this change of the color factor when considering diquarks. Given the parameters of the potential, we obtain the mass of the corresponding diquark in a similar manner as when considering the quark-antiquark system, only changing the string tension $b\to b/2$ and $\kappa\to \kappa/2$, due to the change in the color structure of the system, and thus finding the energy eigenvalues of the diquark systems. 

\subsection{Fitting procedure}

The fitting procedure of the model is described as follows: A fit of the four parameters of the model to experimental data is performed by finding the parameters $\mathbf{v}\equiv(m,\alpha_S,b,\sigma)$, where $m \equiv m_1 = m_2$, that minimizes the function, 
\begin{equation}
    \chi^2 \equiv \sum_{i=1}^{N_{\rm data}} w_i{\left[M_{\text{exp},i}-M_{\text{model},i}(\mathbf{v})\right]^2} \,,
    \label{eq:chi2}
\end{equation}
where $N_{\rm data}$ is the number of experimental data and $M_{\text{exp},i}$ is the experimental mass of the corresponding mass $M_{\text{model},i}(\mathbf{v})$, which is given in the model as a function of $\mathbf{v}$. Each term in Eq.~(\ref{eq:chi2}) is then weighted with $w_i$ for each mass. Following Ref.~\cite{DeSanctis:2010zz}, we will only consider $w_i=1$, giving the same statistical significance to all states used as input. It should be noted that choosing $w_i = 1$, the $\chi^2$ function in Eq.~(\ref{eq:chi2}) will be dimensionful. However, we will choose to present the values of this $\chi^2$ function (as well as individual pulls) without units.

\section{Numerical fits and results}
\label{sec:results}

\subsection{Data sets}

The model will be numerically fit to five different data sets. First, a data set consisting entirely of charmonium mesons (in total 15~mesons). Second, a data set consisting entirely of bottomonium mesons (in total 15~mesons). Third, a data set consisting of $D$ mesons (in total eight~mesons). Fourth, a data set of $B$ mesons (in total seven~mesons). Fifth, a fit to both the charmonium and bottomonium meson data will be made (in total 30~mesons). A meson consisting of two charm quarks is a good candidate to fit to the model, since it has a relatively large constituent mass compared to light quarks, and therefore, a nonrelativistic approach can be justified. Both charmonium and bottomonium are heavy mesons and well suited to the restrictions of this model. For reference, the data set of charmonium mesons is called \textbf{I}, the data set of bottomonium mesons \textbf{II}, the data set consisting of only $D$ mesons \textbf{III}, the data set consisting of only $B$ mesons \textbf{IV}, and finally, the data set containing both charmonium and bottomonium mesons \textbf{V}. In Table~\ref{tab:TotData}, the data used are presented.
\begin{table}
\caption{Meson data adopted from Ref.~\cite{Zyla:2020}. In total, there are 45 mesons and their experimental masses included, which are divided into 15~charmonium meson, 15~bottomonium mesons, 8~$D$ mesons (having a quark content of $c\bar{q}$, where $q=u,d$), and 7~$B$ mesons (having a quark content of $b\bar{q}$, where $q=u,d$). Note that we use a spectroscopic notation, where $N$ denotes the principal quantum number, $S$ the total spin quantum number, $L$ the orbital quantum number, and $J$ the total angular momentum quantum number.}
    \label{tab:TotData}
    \centering
    \begin{tabular}{cccc}
    \hline
    \hline
        Meson &$N^{2S+1}L_J $ & $J^{PC}$/$J^{P}$ & Experimental mass [MeV]\\
        \hline
        $\eta_c(1S)$ & $1^1S_0$ & $0^{-+}$ & $2983.9 \pm 0.5$\\
        $J/\psi(1S)$ & $1^3S_1$ & $1^{--}$ & $3096.900 \pm 0.006$\\
        $\chi_{c0}(1P)$ & $1^3P_0$ & $0^{++}$ & $3414.71 \pm 0.30$\\
        $\chi_{c1}(1P)$ & $1^3P_1$ & $1^{++}$ & $3510.67 \pm 0.05$\\
        $h_c(1P)$ & $1^1P_1$ & $1^{+-}$ & $3525.38 \pm 0.11$\\
        $\chi_{c2}(1P)$ & $1^3P_2$ & $2^{++}$ & $3556.17 \pm 0.07$\\
        $\eta_c(2S)$ & $2^1S_0$ & $0^{-+}$ & $3637.5 \pm 1.1$\\
        $\psi(2S)$ & $2^3S_1$ & $1^{--}$ & $3686.10 \pm 0.06$\\
        $\psi(3770)$ & $1^3D_1$ & $1^{--}$ & $3773.7 \pm 0.4$\\
        $\psi_2(3823)$ & $1^3D_2$ & $2^{--}$ & $3822.2 \pm 1.2$\\
        $\psi_3(3842)$ & $1^3D_3$ & $3^{--}$ & $3842.71 \pm 0.16 \pm 0.12$\\
        $\chi_{c2}(3930)$ & $2^3P_2$ & $2^{++}$ & $3922.2 \pm 1.0$\\
        $\psi(4040)$ & $3^3S_1$ & $1^{--}$ & $4039 \pm 1$\\
        $\psi(4160)$ & $2^3D_1$ & $1^{--}$ & $4191 \pm 5$\\
        $\psi(4415)$ & $4^3S_1$ & $1^{--}$ & $4421 \pm 4$\\
        \hline
        $\eta_b(1S)$ & $1^1S_0$ & $0^{-+}$ & $9398.7 \pm 2.0$\\
        $\Upsilon(1S)$ & $1^3S_1$ & $1^{--}$ & $9460.30 \pm 0.26$\\
        $\chi_{b0}(1P)$ & $1^3P_0$ & $0^{++}$ & $9859.44 \pm 0.42 \pm 0.31$\\
        $\chi_{b1}(1P)$ & $1^3P_1$ & $1^{++}$ & $9892.78 \pm 0.26 \pm 0.31$\\
        $h_b(1P)$ & $1^1P_1$ & $1^{+-}$ & $9899.3 \pm 0.8$\\
        $\chi_{b2}(1P)$ & $1^3P_2$ & $2^{++}$ & $9912.21 \pm 0.26 \pm 0.31$\\
        $\Upsilon(2S)$ & $2^3S_1$ & $1^{--}$ & $10023.26 \pm 0.31$\\
        $\Upsilon_2(1D)$ & $1^3D_2$ & $2^{--}$ & $10163.7 \pm 1.4$\\
        $\chi_{b0}(2P)$ & $2^3P_0$ & $0^{++}$ & $10232.5 \pm 0.4 \pm 0.5$\\
        $\chi_{b1}(2P)$ & $2^3P_1$ & $1^{++}$ & $10255.46 \pm 0.22 \pm 0.50$\\
        $\chi_{b2}(2P)$ & $2^3P_2$ & $2^{++}$ & $10268.65 \pm 0.22 \pm 0.50$\\
        $\Upsilon(3S)$ & $3^3S_1$ & $1^{--}$ & $10355.2 \pm 0.5$\\
        $\chi_{b1}(3P)$ & $3^3P_1$ & $1^{++}$ & $10513.42 \pm 0.41 \pm 0.53$\\
        $\chi_{b2}(3P)$ & $3^3P_2$ & $2^{++}$ & $10524.02 \pm 0.57 \pm 0.53$\\
        $\Upsilon(4S)$ & $4^3S_1$ & $1^{--}$ & $10579.4 \pm 1.2$\\
        \hline 
        $D^0$ & $1^1S_0$ & $0^-$ & $1864.83 \pm 0.05$\\
        $D^\pm$ & $1^1S_0$ & $0^-$ & $1869.65 \pm 0.05$\\
        $D^{*}(2007)^{0}$ & $1^3S_1$ & $1^-$ & $2006.85 \pm 0.05$\\
        $D^{*}(2010)^\pm$ & $1^3S_1$ & $1^-$ & $2010.26 \pm 0.05$\\
        $D^*_0(2300)^0$ & $1^3P_0 $ & $0^+$ & $2300 \pm 19$\\
        $D_1(2420)^0$ & $1^1P_1$ & $1^+$ & $2420.8 \pm 0.5$\\
        $D^*_2(2460)^0$ & $1^3P_2$ & $2^+$ & $2460.7 \pm 0.4$\\
        $D^*_2(2460)^\pm$ & $1^3P_2$ & $2^+$ & $2465.4 \pm 1.3$\\
        \hline
        $B^\pm$ & $1^1S_0$ & $0^-$ & $5279.34 \pm 0.12$\\
        $B^0 $ & $1^1S_0$ & $0^-$ & $5279.65 \pm 0.12$\\
        $B^*$ & $1^3S_1$ & $1^-$ & $5324.70 \pm 0.21$\\
        $B_1(5721)^+$ & $1^1P_1$ & $1^+$ & $5725.9^{+2.5}_{-2.7}$\\
        $B_1(5721)^0$ & $1^1P_1$ & $1^+$ & $5726.1 \pm 1.3$\\
        $B_2^*(5747)^+$ & $1^3P_2$ & $2^+$ & $5737.2 \pm 0.7$\\
        $B_2^*(5747)^0$ & $1^3P_2$ & $2^+$ & $5739.5 \pm 0.7$\\
        \hline
        \hline
    \end{tabular}
\end{table}

\subsection{Numerical fits and results}

In this subsection, the results of the fitted data sets, and subsequently, the resulting masses of different diquarks and tetraquarks are presented. The procedure can be divided into three main parts. First, fitting the model to each data set~\textbf{I}--\textbf{V} to obtain five sets of parameter values for the free parameters of the model. Next, using the sets of parameter values obtained by fitting data sets~\textbf{I}--\textbf{IV} to calculate the masses of different diquarks. In detail, the sets of parameter values obtained by fitting data sets~\textbf{I}, \textbf{II}, \textbf{III}, and \textbf{IV} are used to calculate the masses of the $cc$, $bb$, $qc$, and $qb$ diquarks, respectively, with $q$ being either an up quark ($u$) or a down quark ($d$). Finally, using the calculated diquark masses to calculate the masses of different tetraquarks. The set of parameter values used for this computation is the one obtained by fitting data set~\textbf{V} to the model. 

The number of free parameters when fitting the model to data sets~\textbf{I} and \textbf{II} is four, since the masses of the constituent quarks for those data sets are equal, i.e., $m = m_1 = m_2$. When fitting the model to data sets~\textbf{III}--\textbf{V}, we use the values for the constituent masses of the charm and bottom quarks obtained in the fits to data sets~\textbf{I} and \textbf{II}, which means that the number of free parameters is three. Also, when considering data sets~\textbf{III} and \textbf{IV}, we use the value $0.323$ GeV as the constituent mass of an up quark or a down quark, which is taken from Ref.~\cite{Griffiths:111880} (see also p.~1 in the review ``59.~Quark Masses'' \cite{Zyla:2020}). 

In practice, we are solving Eq.~(\ref{eq:SchrodingerRed2}) in the eigenbasis of the spin operators $\mathbf{S}$, $\mathbf{S}_1$, and $\mathbf{S}_2$, thus effectively replacing the product $\mathbf{S}_1\cdot\mathbf{S}_2$ by
\begin{equation}
\mathbf{S}_1\cdot\mathbf{S}_2 = \frac{1}{2}\left[S(S+1)-S_1(S_1+1)-S_2(S_2+1)\right] \,,
\label{eq:spin4}
\end{equation}
where $S$, $S_1$, and $S_2$ is the total spin, the spin of quark~1, and the spin of quark~2, respectively. However, note that this modeling procedure is able to split the masses of states with equal principal ($N$), orbital ($L$), and spin ($S$) quantum numbers, but not different total angular momentum ($J$) quantum numbers, i.e., the model is independent of $J$. Solving the Schr{\"o}dinger equation numerically is performed by assuming Dirichlet boundary conditions at $r=0$ and $r=r_{0}$ when using Eq.~(\ref{eq:SchrodingerRed2}). The value of the parameter $r_0$ is chosen so that the energy eigenvalue $E$ is independent of $r_0$ up to five significant digits. This approach was inspired by the method described in Ref.~\cite{Lucha:1998xc}.

Next, the minimization of Eq.~(\ref{eq:chi2}) is initially carried out by performing a random search with $n = 100~000$~points in the parameter space spanned by the parameters $\mathbf{v} = (m,\alpha_S,b,\sigma)$. The conditions on the parameters are chosen to be
\begin{align}
0.05 \leq \; &\alpha_S \leq 0.70 \,, \nonumber\\
0.01~{\rm GeV}^2 \leq \; &b \leq 0.40~{\rm GeV}^2 \,, \nonumber\\
0.05~{\rm GeV} \leq \; &\sigma \leq 1.50~{\rm GeV} \,, \nonumber
\end{align}
as well as $1.00~{\rm GeV} \leq m \leq 2.00~{\rm GeV}$ for data set~\textbf{I} and $4.00~{\rm GeV} \leq m \leq 5.00~{\rm GeV}$ for data set~\textbf{II}. Furthermore, for data sets~\textbf{III}--\textbf{V}, the values of $m$ obtained for data sets~\textbf{I} and \textbf{II} are used as input values. After the initial random search, an iterative adaptive method, using the same technique but with narrower conditions on the parameters and significantly smaller number of points, is performed to optimize the coarse point found during the initial random search in order to obtain the (local) best-fit point that minimizes the $\chi^2$ function in Eq.~(\ref{eq:chi2}). In Table~\ref{tab:Pulls}, the resulting values of the $\chi^2$ function for the five data sets~\textbf{I}--\textbf{V} are presented as well as each pull of each meson is listed, and in Table~\ref{tab:ParametersFIt}, the resulting parameter values for $m$, $\alpha_S$, $b$, and $\sigma$ when fitting the model to the respective data sets are given. 
\begin{table*}
\caption{Resulting value of the $\chi^2$ function for each fit to data sets~\textbf{I}--\textbf{V} to the model with respective pull for each meson data point.}
    \label{tab:Pulls}
    \centering
    \begin{tabular}{cccccc}
    \hline
    \hline
        Meson & Data set~\textbf{I} & Data set~\textbf{II} & Data set~\textbf{III} & Data set~\textbf{IV} & Data set~\textbf{V}\\
        \hline
        $\eta_c(1S)$ & $-4.1 \times 10^{-3}$ & $\cdots$ & $\cdots$ & $\cdots$ & $-1.6 \times 10^{-1}$\\
        $J/\psi(1S)$ & $1.2 \times 10^{-2}$ & $\cdots$ & $\cdots$ & $\cdots$ & $-1.2 \times 10^{-1}$\\
        $\chi_{c0}(1P)$ & $-1.0 \times 10^{-1}$ & $\cdots$ & $\cdots$ & $\cdots$ & $-1.6 \times 10^{-1}$\\
        $\chi_{c1}(1P)$ & $-6.1 \times 10^{-3}$ & $\cdots$ & $\cdots$ & $\cdots$ & $-6.4 \times 10^{-2}$\\
        $h_c(1P)$ & $1.8 \times 10^{-2}$&$\cdots$ & $\cdots$ & $\cdots$ & $-4.6 \times 10^{-2}$\\
        $\chi_{c2}(1P)$ & $3.9 \times 10^{-2}$ & $\cdots$ & $\cdots$ & $\cdots$ & $-1.8 \times 10^{-2}$\\  
        $\eta_c(2S)$ & $5.8 \times 10^{-3}$ & $\cdots$ & $\cdots$ & $\cdots$ & $-5.9 \times 10^{-2}$\\ 
        $\psi(2S)$ & $1.6 \times 10^{-2}$ & $\cdots$ & $\cdots$ & $\cdots$ & $-5.0 \times 10^{-2}$\\ 
        $\psi(3770)$ & $-2.2 \times 10^{-2}$ & $\cdots$ & $\cdots$ & $\cdots$ & $-5.7 \times 10^{-2}$\\
        $\psi_2(3823)$ & $2.6 \times 10^{-2}$ & $\cdots$ & $\cdots$ & $\cdots$ & $-8.4 \times 10^{-3}$\\
        $\psi_3(3842)$ & $4.7 \times 10^{-2}$ & $\cdots$ & $\cdots$ & $\cdots$ & $1.2 \times 10^{-2}$\\
        $\chi_{c2}(3930)$ & $-2.2 \times 10^{-2}$ & $\cdots$ & $\cdots$ & $\cdots$ & $-5.6 \times 10^{-2}$\\
        $\psi(4040)$ & $-3.9 \times 10^{-2}$ & $\cdots$ & $\cdots$ & $\cdots$ & $-8.1 \times 10^{-2}$ \\
        $\psi(4160)$ & $2.7 \times 10^{-2}$ & $\cdots$ & $\cdots$ & $\cdots$ & $6.7 \times 10^{-3}$\\
        $\psi(4415)$ & $9.4 \times 10^{-4}$ & $\cdots$ & $\cdots$ & $\cdots$ & $-2.7 \times 10^{-2}$\\
        \hline
        $\eta_b(1S)$ & $\cdots$ & $-1.3 \times 10^{-2}$ & $\cdots$ & $\cdots$ & $-1.4 \times 10^{-2}$\\
        $\Upsilon(1S)$ & $\cdots$ & $1.7 \times 10^{-2}$ & $\cdots$& $\cdots$ & $1.9 \times 10^{-2}$\\
        $\chi_{b0}(1P)$ & $\cdots$ & $-4.5 \times 10^{-2}$ & $\cdots$ & $\cdots$ & $-3.1 \times 10^{-3}$\\
        $\chi_{b1}(1P)$ & $\cdots$ & $-1.2 \times 10^{-2}$ & $\cdots$ & $\cdots$ & $3.0 \times 10^{-2}$\\
        $h_b(1P)$ & $\cdots$ & $-1.8 \times 10^{-3}$ & $\cdots$ & $\cdots$ & $4.0 \times 10^{-2}$\\
        $\chi_{b2}(1P)$ & $\cdots$ & $7.7 \times 10^{-3}$ & $\cdots$ & $\cdots$ & $5.0 \times 10^{-2}$\\
        $\Upsilon(2S)$ & $\cdots$ & $2.4 \times 10^{-2}$ & $\cdots$ & $\cdots$ & $7.6 \times 10^{-2}$\\
        $\Upsilon_2(1D)$ & $\cdots$ & $1.9 \times 10^{-2}$ & $\cdots$ & $\cdots$ & $8.5 \times 10^{-2}$\\
        $\chi_{b0}(2P)$ & $\cdots$ & $-1.6 \times 10^{-2}$ & $\cdots$ & $\cdots$ & $6.1 \times 10^{-2}$\\
        $\chi_{b1}(2P)$ & $\cdots$ & $7.3 \times 10^{-3}$ & $\cdots$ & $\cdots$ & $8.4 \times 10^{-2}$\\
        $\chi_{b2}(2P)$ & $\cdots$ & $2.1 \times 10^{-2}$ & $\cdots$ & $\cdots$ & $9.7 \times 10^{-2}$\\
        $\Upsilon(3S)$ & $\cdots$ & $2.2 \times 10^{-2}$ & $\cdots$ & $\cdots$ & $1.1 \times 10^{-1}$\\
        $\chi_{b1}(3P)$ & $\cdots$ & $-1.0 \times 10^{-2}$ & $\cdots$ & $\cdots$ & $9.5 \times 10^{-2}$\\
        $\chi_{b2}(3P)$ & $\cdots$ & $3.5 \times 10^{-4}$ & $\cdots$ & $\cdots$ & $1.1 \times 10^{-1}$\\
        $\Upsilon(4S)$ & $\cdots$ & $-2.3 \times 10^{-2}$ & $\cdots$ & $\cdots$ & $9.0 \times 10^{-2}$\\
        \hline 
        $D^0$ & $\cdots$ & $\cdots$ & $-7.1 \times 10^{-3}$ & $\cdots$ & $\cdots$\\
        $D^+$ & $\cdots$ & $\cdots$ & $-2.3 \times 10^{-3}$ & $\cdots$ & $\cdots$\\
        $D^{*}(2007)^0$ & $\cdots$ & $\cdots$ & $-3.8 \times 10^{-2}$ & $\cdots$ & $\cdots$\\
        $D^{*}(2010)^\pm$ & $\cdots$ & $\cdots$ & $-3.5 \times 10^{-2}$ & $\cdots$ & $\cdots$\\
        $D^{*}_0(2300)^0$ & $\cdots$ & $\cdots$ & $-1.0 \times 10^{-1}$ & $\cdots$ & $\cdots$\\
        $D_1(2420)^0$ & $\cdots$ & $\cdots$ & $3.6 \times 10^{-2}$ & $\cdots$ & $\cdots$\\
        $D^{*}_2(2460)^0$ & $\cdots$ & $\cdots$ & $5.7 \times 10^{-2}$ & $\cdots$ & $\cdots$\\
        $D^{*}_2(2460)^\pm$ & $\cdots$ & $\cdots$ & $6.2 \times 10^{-2}$ & $\cdots$ & $\cdots$\\
        \hline
        $B^\pm$ & $\cdots$ & $\cdots$ & $\cdots$ & $-2.4 \times 10^{-3}$ & $\cdots$\\
        $B^0$ & $\cdots$ & $\cdots$ & $\cdots$ & $-2.1 \times 10^{-3}$ & $\cdots$\\
        $B^*$ & $\cdots$ & $\cdots$ & $\cdots$ & $-9.7 \times 10^{-3}$ & $\cdots$\\
        $B_1(5721)^+$ & $\cdots$ & $\cdots$ & $\cdots$ & $5.3 \times 10^{-3}$ & $\cdots$\\
        $B_1(5721)^0$ & $\cdots$ & $\cdots$ & $\cdots$ & $5.5 \times 10^{-3}$ & $\cdots$\\ 
        $B^*_2(5747)^+$ & $\cdots$ & $\cdots$ & $\cdots$ & $-2.3 \times 10^{-3}$ & $\cdots$\\
        $B^*_2(5747)^0$ & $\cdots$ & $\cdots$ & $\cdots$ & $-3.2 \times 10^{-5}$ & $\cdots$\\
        \hline 
        $\chi^2$ & $1.9 \times 10^{-2}$ & $5.5 \times 10^{-3}$ & $2.2 \times 10^{-2}$ & $1.7 \times 10^{-4}$ & $1.7 \times 10^{-1}$\\
        \hline
        \hline
    \end{tabular}
\end{table*}
\begin{table}[ht]
\caption{Resulting best-fit parameter values for $m$, $\alpha_S$, $b$, and $\sigma$ from each fit of data sets~\textbf{I}--\textbf{V} to the model (including also the corresponding value of the $\chi^2$ function). Note that there are no resulting values for $m$ for the fits to data sets~\textbf{III}--\textbf{V}. For data set~\textbf{III}, the value for the charm quark mass 1.459~GeV from data set~\textbf{I} is used as an input value, whereas for data set~\textbf{IV}, the value for the bottom quark mass 4.783~GeV from data set~\textbf{II} is used as an input value. For both data sets~\textbf{III} and \textbf{IV}, the value for a light quark mass~0.323~GeV is used \cite{Griffiths:111880}. For data set~\textbf{V}, the values for both the charm and bottom quarks from data sets~\textbf{I} and \textbf{II}, respectively, are used as input values.}
    \label{tab:ParametersFIt}
    \centering
    \begin{tabular}{cccccc}
    \hline
    \hline
        Data set & $m~[{\rm GeV}]$& $\alpha_S$ & $b~[{\rm GeV}^2]$ & $\sigma~[{\rm GeV}]$ & $\chi^2$\\
        \hline
        \textbf{I} & 1.459 & 0.5234 & 0.1480 & 1.048 & 0.01887\\
        \textbf{II} & 4.783 & 0.3841 & 0.1708 & 1.50 & 0.005485\\
        \textbf{III} & $\cdots$ & 0.70 & 0.08417 & 0.5760 & 0.02182\\
        \textbf{IV} & $\cdots$ & 0.70 & 0.09467 & 0.3049 & 0.0001696\\
        \textbf{V} & $\cdots$ & 0.3714 & 0.1445 & 1.50 & 0.1694\\
        \hline
        \hline
    \end{tabular}
\end{table}

\subsubsection{Diquarks}
\label{subsub:diquarks}

Given the best-fit values for the free parameters of the model, we find the diquark masses by calculating the energy eigenvalues, changing $\kappa \to \kappa/2$ and $b\to b/2$ in order to compensate for the change in color structure of the quark-quark system (compared to the color structure of the quark-antiquark system, see the discussion in Sec.~\ref{sub:color}). It should also be noted that when we change from the quark-antiquark system (i.e., mesons) to the quark-quark system (i.e., diquarks) the information on the spin-spin interaction [described by Eqs.~(\ref{eq:spin1}), (\ref{eq:spin2}), (\ref{eq:spin3}), and (\ref{eq:spin4})] is inherited in terms of the free parameters, and especially, the parameter $\sigma$. Therefore, the spin-spin interaction is encoded in the diquarks. Note that the idea that the spin-spin interaction within each diquark being the most relevant one was proposed in Ref.~\cite{Maiani:2014aja}. The sets of parameter values obtained when fitting the model to data sets~\textbf{I}, \textbf{II}, \textbf{III}, and \textbf{IV} are used to calculate the masses of the $cc$, $bb$, $qc$, and $qb$ diquarks, respectively. We consider only diquarks in the ground state $N^{2S+1}L_J=1^3S_1$, which are known as axial-vector diquarks and named \emph{good diquarks} by Jaffe \cite{Jaffe:2004ph}. In Table~\ref{tab:DiquarkResults}, the results for the four diquark masses are presented.
\begin{table}[ht]
\caption{Results for the four diquark masses $M$ and the corresponding energy eigenvalues $E$ for each of the diquarks, where $q$ represents either an up or a down quark. All diquarks are considered to be in the ground state $N^{2S+1}L_J=1^3S_1$.}
    \label{tab:DiquarkResults}
    \centering
    \begin{tabular}{cccc}
        \hline
        \hline 
        Diquark & Data set & $E~[{\rm MeV}]$ & $M~[{\rm MeV}]$\\
        \hline 
        $qc$ & \textbf{III} & 236.6 & 2018\\
        $cc$ & \textbf{I} & 210.6 & 3128\\
        $qb$ & \textbf{IV} & 232.8 & 5339\\
        $bb$ & \textbf{II} & 76.40 & 9643\\
        \hline
        \hline
    \end{tabular}
\end{table}

\subsubsection{Tetraquarks}

For tetraquarks, we consider them to be composites of (axial-vector) diquarks and antidiquarks and the interaction between the diquarks and the antidiquarks is assumed to be effectively the same as for ordinary quarkonia. Thus, the parameter set obtained when fitting data set~\textbf{V} to the model is used in the effective potential for all tetraquarks. However, when considering $cc\overline{cc}$ and $bb\overline{bb}$ tetraquarks, we also compute the tetraquark masses with the parameter sets found by fitting the model to data sets~\textbf{I} and \textbf{II}, respectively. Since the diquarks and antidiquark are in the antitriplet and triplet color states, respectively, the color structure of tetraquarks has identical color structure as the mesons, and subsequently, the same color factor $\kappa=-4/3$. In addition, the same string tension $b$ is used for tetraquarks as for the mesons. Concerning the spin-spin interaction, it is passed on to tetraquarks in a similar way as it is carried over from the mesons to the diquarks, as described in Sec.~\ref{subsub:diquarks}. In Refs.~\cite{Maiani:2017kyi,Esposito:2018cwh}, based on the spin-spin interaction, the hypothesis of a separation of distance scales between diquarks and antidiquarks in tetraquarks was introduced by a potential barrier. Especially, in Ref.~\cite{Esposito:2018cwh}, the tetraquark potential modeled by a repulsive barrier is explicitly given (not considering the Coulomb term) by $V(r) \propto b (r-\ell)$, where $\ell$ is the width of the barrier, which effectively changes the parameter $b$. In our model for tetraquarks, the potential includes the same parts as for mesons and diquarks; i.e., it is described by the parameters $\kappa$, $\alpha_S$, $b$, $\sigma$, and $L$, although with different values of $\kappa$ and $b$, which are both parameters in the Cornell potential [see Eq.~(\ref{eq:Cornell})], than for diquarks but with the same as for mesons. Therefore, the potential effectively changes when considering tetraquarks instead of diquarks. In comparison, both our model and the model introduced in Ref.~\cite{Esposito:2018cwh} effectively change the string tension $b$ when considering tetraquarks. Nevertheless, in our model for tetraquarks, the main effect of the spin-spin interaction stems from the interaction between diquarks and antidiquarks, whereas in the model presented in Ref.~\cite{Maiani:2014aja} for tetraquarks, it is argued that the spin-spin interaction is dominated by this interaction inside diquarks and antidiquarks. Now, in our model, considering diquarks and antidiquarks as consituents of tetraquarks, the tetraquark masses can be calculated using the diquark masses in Table~\ref{tab:DiquarkResults}. In Table~\ref{tab:TetraquarkResults}, the results for the masses of 24 tetraquark states are presented.
\begin{table}[ht]
\caption{Results for masses of 24 tetraquark states $M$ and the corresponding energy eigenvalues $E$ for each of the tetraquark states. The constituent diquarks are assumed to be in the ground state $N^{2S+1}L_J=1^3S_1$, which are to be found in Table~\ref{tab:DiquarkResults}. The label ``Data sets ${\bf n}_1+{\bf n}_2$'' indicates that the input values for the diquark masses are adopted from data set~${\bf n}_1$, and the input values for the other parameters are taken from data set~${\bf n}_2$.}
    \label{tab:TetraquarkResults}
    \centering
    \begin{tabular}{ccccccc}
    \hline
    \hline 
    Tetraquark & $N^{2S+1}L$ & $E~[{\rm MeV}]$ & $M~[{\rm MeV}]$ & $E~[{\rm MeV}]$ & $M~[{\rm MeV}]$\\
                          \hline
$qc\overline{qc}$ & & & & \multicolumn{2}{c}{{\footnotesize Data sets~\textbf{III+V}}}\\
                          & $1^1S$ & & & 39.63 & 4076\\
                          & $1^3S$ & & & 119.4 & 4156\\
                          & $1^5S$ & & & 225.6 & 4262\\
                          & $1^1P$ & & & 545.4 & 4582\\
                          & $1^3P$ & & & 549.0 & 4585\\
                          & $1^5P$ & & & 555.2 & 4591\\
                          & $1^1D$ & & & 792.9 & 4829\\
                          & $1^3D$ & & & 793.1 & 4829\\
                          & $1^5D$ & & & 793.5 & 4830\\
                          \hline
$cc\overline{cc}$ & & \multicolumn{2}{c}{{\footnotesize Data sets~\textbf{I+I}}} & \multicolumn{2}{c}{{\footnotesize Data sets~\textbf{I+V}}}\\                          
                          & $1^1S$ & $-295.5$ & 5960 & $-56.91$ & 6198 \\
                          & $1^3S$ & $-246.4$ & 6009 & $-9.427$ & 6246\\
                          & $1^5S$ & $-155.6$ & 6100 & 67.94 & 6323\\
                          \hline
$qb\overline{qb}$ & & & & \multicolumn{2}{c}{{\footnotesize Data sets~\textbf{IV+V}}}\\  
                          & $1^1S$ & & & $-234.0$ & 10445\\
                          & $1^3S$ & & & $-206.6$ & 10472\\
                          & $1^5S$ & & & $-155.7$ & 10523\\
                          & $1^1P$ & & & 257.3 & 10936\\
                          & $1^3P$ & & & 260.2 & 10939\\
                          & $1^5P$ & & & 265.8 & 10944\\
                          & $1^1D$ & & & 478.8 & 11157\\
                          & $1^3D$ & & & 479.0 & 11158\\
                          & $1^5D$ & & & 479.6 & 11158\\
                          \hline
$bb\overline{bb}$ & & \multicolumn{2}{c}{{\footnotesize Data sets~\textbf{II+II}}} & \multicolumn{2}{c}{{\footnotesize Data sets~\textbf{II+V}}}\\ 
                          & $1^1S$ & $-563.5$ & 18723 & $-532.3$ & 18754\\
                          & $1^3S$ & $-548.2$ & 18738 & $-518.0$ & 18768\\
                          & $1^5S$ & $-518.1$ & 18768 & $-489.9$ & 18797\\
                          \hline
                          \hline 
    \end{tabular}
\end{table}

\subsection{Motivation of parameters and comparison with other works}

Similar models to the model presented in this work have been proposed in Refs.~\cite{Debastiani:2017msn, Debastiani:2015fea}. In Ref.~\cite{Debastiani:2017msn}, the authors were using the same model as in this work, but also taking into account perturbation in the spin-orbit and tensor interactions, although considering only fully charmed diquarks and tetraquarks (i.e., $cc$ and $cc\overline{cc}$), and thus, their results can be compared to the results for the set of parameters fitted to data set~\textbf{I}. The models are identical for those states, where the perturbation energy is zero. In Ref.~\cite{Debastiani:2015fea}, the same authors considered $X(3872)$ [also known as $\chi_{c1}(3872)$ \cite{Zyla:2020}] under the hypothesis that its constituents are consisting of a diquark $qc$ and an antidiquark $\overline{qc}$, and their parameter values could therefore be compared to the parameter values found by fitting data set~\textbf{III}. They fit the model in order to investigate if $Z_c(4430)$ could be an exited state of $X(3872)$. In Ref.~\cite{Maiani:2014aja}, the influence of the spin-spin interaction, where this interaction between $q$ and $c$ quarks inside a diquark dominates compared to all other possibilities, was introduced and showed to be able to predict the main features of the observed spectra of  tetraquarks. Furthermore, we will compare the masses of diquarks and tetraquarks calculated in this model with those presented in Refs.~\cite{Bedolla:2019zwg, Maiani:2004vq, Debastiani:2017msn, Bedolla:2019noq, Esau:2019hqw, Liu:2019zuc, Patel:2014vua, Ebert:2005nc,Lu:2016zhe, Anwar:2017toa, Berezhnoy:2011xn, Karliner:2016zzc, Barnea:2006sd, Chen:2020lgj, Hadizadeh:2015cvx, Wu:2016vtq, Wang:2019rdo}. In Tables~\ref{tab:Paramters1Lit}, \ref{tab:DiquarkLit}, and \ref{tab:TetraquarkResultsComparison}, we display the different comparisons.
\begin{table}[ht]
\caption{Comparison of masses $M$ and model parameters $\alpha_S$, $b$, and $\sigma$ for the effective potential (including spin-spin interaction) between this work and two other works. An asterisk (``*'') indicates that the value is an input value.}
    \label{tab:Paramters1Lit}
    \centering
    \begin{tabular}{ccccc}
       \hline
       \hline
       Source & $M~[{\rm GeV}]$ & $\alpha_S$ & $b~[{\rm GeV}^2]$ & $\sigma~[{\rm GeV}]$\\
       \hline
       \textbf{I} (PDG 2020) & 1.459 & 0.5234 & 0.1480 & 1.048\\
       \textbf{I} (PDG 2017) & 1.442 & 0.4951 & 0.1501 & 1.150\\
       Ref.~\cite{Debastiani:2017msn}& 1.4622 & 0.5202 & 0.1463 & 1.0831\\
       \hline
       \textbf{III} (PDG 2020) & $1.459^*$ & 0.70 & 0.08417 & 0.5760\\
       Refs.~\cite{Patel:2014vua, Debastiani:2015fea} & $1.486^*$ & 0.30 & 0.015 & $\cdots$\\
       \hline
       \hline
    \end{tabular}
\end{table}
\begin{table*}[ht]
\caption{Comparison of masses of diquarks, in units of MeV, between the masses $M$ calculated in this work and values from other different works. All diquarks are considered to be in the ground state $N^{2S+1}L_J=1^3S_1$.}
    \label{tab:DiquarkLit}
    \centering
    \begin{tabular}{cccccccccccc}
       \hline
       \hline
       Diquark & $M~[{\rm MeV}]$ & Ref.~\cite{Bedolla:2019noq} & Ref.~\cite{Maiani:2004vq} & Ref.~\cite{Debastiani:2017msn} & Ref.~\cite{Ebert:2005nc} & Ref.~\cite{Lu:2016zhe}  & Ref.~\cite{Wang:2010sh} & Refs.~\cite{Kleiv:2013dta, Esau:2019hqw} & Refs.~\cite{Anwar:2017toa, Anwar:2018sol} & Ref.~\cite{Berezhnoy:2011xn} & Ref.~\cite{Karliner:2016zzc}\\
       \hline 
       $qc$ & 2018 & 2138 & 1933 & $\cdots$ & 2036 & 2138 & $1760 \pm 80$ & $1870 \pm 100$ & 2250 & $\cdots$ & $\cdots$\\
       $cc$ & 3128 & 3329 & $\cdots$ & 3133.4 & $\cdots$ & $\cdots$ & $\cdots$ & $3510 \pm 350$ & $\cdots$ & 3130 & 3204.1\\
       $qb$ & 5339 & 5465 & $\cdots$ & $\cdots$ & 5381 & 5465 & $5130 \pm 110$ & $5080 \pm 40$ & $\cdots$ & $\cdots$ & $\cdots$\\
       $bb$ & 9643 & 9845 & $\cdots$ & $\cdots$ & $\cdots$ & $\cdots$ & $\cdots$ & $8670 \pm 690$ & 9850 & 9720 & 9718.9\\
       \hline
       \hline
    \end{tabular}
\end{table*}
\begin{table*}[ht]
\caption{Comparison of masses of tetraquarks, in units of MeV, between the masses $M$ calculated in this work and values from other different works investigating tetraquarks. For the $cc\overline{cc}$ and $bb\overline{bb}$ tetraquark masses, the upper values correspond to data sets~\textbf{I+I} and \textbf{II+II}, respectively, whereas the lower values correspond to data sets~\textbf{I+V} and \textbf{II+V}, respectively.}
    \label{tab:TetraquarkResultsComparison}
    \centering
    \begin{tabular}{ccccccccccccccccc}
    \hline
    \hline 
    $\begin{array}{c} \mbox{Tetra-} \\ \mbox{quark} \end{array}$ & $N^{2S+1}L$ & $(J^{PC})$ & $M~[{\rm MeV}]$ & $\begin{array}{c} \mbox{Ref.} \\ \mbox{\cite{Patel:2014vua}} \end{array}$ & $\begin{array}{c} \mbox{Ref.} \\ \mbox{\cite{Barnea:2006sd}} \end{array}$ & $\begin{array}{c} \mbox{Ref.} \\ \mbox{\cite{Debastiani:2017msn}} \end{array}$ & $\begin{array}{c} \mbox{Ref.} \\ \mbox{\cite{Chen:2020lgj}} \end{array}$ & $\begin{array}{c} \mbox{Ref.} \\ \mbox{\cite{Ebert:2005nc}} \end{array}$ & $\begin{array}{c} \mbox{Ref.} \\ \mbox{\cite{Hadizadeh:2015cvx}} \end{array}$ & $\begin{array}{c} \mbox{Refs.} \\ \mbox{\cite{Anwar:2017toa, Anwar:2018sol}} \end{array}$ & $\begin{array}{c} \mbox{Ref.} \\ \mbox{\cite{Bedolla:2019zwg}} \end{array}$ & $\begin{array}{c} \mbox{Ref.} \\ \mbox{\cite{Berezhnoy:2011xn}} \end{array}$ & $\begin{array}{c} \mbox{Ref.} \\ \mbox{\cite{Wu:2016vtq}} \end{array}$ & $\begin{array}{c} \mbox{Ref.} \\ \mbox{\cite{Liu:2019zuc}} \end{array}$ & $\begin{array}{c} \mbox{Ref.} \\ \mbox{\cite{Wang:2019rdo}} \end{array}$ & $\begin{array}{c} \mbox{Ref.} \\ \mbox{\cite{Wang:2018poa}} \end{array}$\\
                          \hline
                          & $1^1S$ & $(0^{++})$ & 4076 & 3849 & $\cdots$ & $\cdots$ & $\cdots$ & 3852 & $\begin{array}{c} 3919 \\ 3869 \end{array}$ & 3641 & $\cdots$ & $\cdots$ & $\cdots$ & $\cdots$ & $\cdots$ & $\cdots$\\
        $qc\overline{qc}$ & $1^3S$ & $(1^{+-})$ & 4156 & 3822 & $\cdots$ & $\cdots$ & $\cdots$ & 3890 & $\cdots$ & 4047 & $\cdots$ & $\cdots$ & $\cdots$ & $\cdots$ & $\cdots$ & $\cdots$\\
                          & $1^5S$ & $(2^{++})$ & 4262 & 3946 & $\cdots$ & $\cdots$ & $\cdots$ & 3968 & $\cdots$ & $\cdots$ & $\cdots$ & $\cdots$ & $\cdots$ & $\cdots$ & $\cdots$ & $\cdots$\\
                          \hline 
                          & $1^1S$ & $(0^{++})$ & $\begin{array}{c} 5960 \\ 6198 \end{array}$ & $\cdots$ & 6038 & 5969.4 & 6360.2 & $\cdots$ & $\cdots$ & $\cdots$ & 5883 & 5966 & 7016 & 6487 & $\begin{array}{c} 6420 \\ 6436 \end{array}$ & 5990\\
        $cc\overline{cc}$ & $1^3S$ & $(1^{+-})$ & $\begin{array}{c} 6009 \\ 6246 \end{array}$ & $\cdots$ & 6101 & 6020.9 & 6397.6 & $\cdots$ & $\cdots$ & $\cdots$ & 6120 & 6051 & 6899 & 6500 & $\begin{array}{c} 6425 \\ 6450 \end{array}$ & 6050\\
                          & $1^5S$ & $(2^{++})$ & $\begin{array}{c} 6100 \\ 6323 \end{array}$ & $\cdots$ & 6172 & $6115.4$ & 6410.4 & $\cdots$ & $\cdots$ & $\cdots$ & 6246 & 6223 & 6956 & 6524 & $\begin{array}{c} 6432 \\ 6479 \end{array}$ & 6090\\
                          \hline 
                          & $1^1S$ & $(0^{++})$ & 10445 & $\cdots$ & $\cdots$ & $\cdots$ & $\cdots$ & 10473 & $\begin{array}{c} 10469 \\ 10453 \end{array}$ & $\cdots$ & 10120 & $\cdots$ & $\cdots$ & $\cdots$ & $\cdots$ & $\cdots$\\
        $qb\overline{qb}$ & $1^3S$ & $(1^{+-})$ &10472 & $\cdots$ & $\cdots$ & $\cdots$ & $\cdots$ & 10494 & $\cdots$ & $\cdots$ & $\cdots$ & $\cdots$ & $\cdots$ & $\cdots$ & $\cdots$ & $\cdots$\\
                          & $1^5S$ & $(2^{++})$ & 10523 & $\cdots$ & $\cdots$ & $\cdots$ & $\cdots$ & 10534 & $\cdots$ & $\cdots$ & $\cdots$ & $\cdots$ & $\cdots$ & $\cdots$ & $\cdots$ & $\cdots$\\
                          \hline 
                          & $1^1S$ & $(0^{++})$ & $\begin{array}{c} 18723 \\ 18754 \end{array}$ & $\cdots$ & $\cdots$ & $\cdots$ & $\cdots$ & $\cdots$ & $\cdots$ & 18750 & 18748 & 18754 & 20275 & 19322 & $\begin{array}{c} 19246 \\ 19297 \end{array}$ & 18840\\
        $bb\overline{bb}$ & $1^3S$ & $(1^{+-})$ & $\begin{array}{c} 18738 \\ 18768 \end{array}$ & $\cdots$ & $\cdots$ & $\cdots$ & $\cdots$ & $\cdots$ & $\cdots$ & $\cdots$ & 18828 & 18808 & 20212 & 19329 & $\begin{array}{c} 19247 \\ 19311 \end{array}$ & 18840\\
                          & $1^5S$ & $(2^{++})$ & $\begin{array}{c} 18768 \\ 18797 \end{array}$ & $\cdots$ & $\cdots$ & $\cdots$ & $\cdots$ & $\cdots$ & $\cdots$ & $\cdots$ & 18900 & 18916 & 20243 & 19341 & $\begin{array}{c} 19249 \\ 19325 \end{array}$ & 18850\\
                          \hline
                          \hline 
    \end{tabular}
\end{table*}

\subsection{Discussion on results}

A thorough discussion on the results obtained in this work is in order. In Table~\ref{tab:Pulls}, the pulls and the values of the $\chi^2$ function from the five fits to data sets~\textbf{I}--\textbf{V} are presented. Comparing the values of the $\chi^2$ function among the fits, we observe that the value for data set~\textbf{V} is the largest with an order of magnitude of $10^{-1}$, the values for data sets~\textbf{I} and \textbf{III} with orders of magnitude of $10^{-2}$, the value for data set~\textbf{II} with an order of magnitude $10^{-3}$, and finally, the value for data set~\textbf{IV} is the smallest with an order of magnitude of $10^{-4}$. The discrepancy in the values of the $\chi^2$ function between data sets~\textbf{IV} and \textbf{V} could be a consequence of the much larger variation of the masses in data set~\textbf{V} compared to the variation of the masses in data set~\textbf{IV}. Also, one could expect that, when fitting the model to data set~\textbf{V}, the value of the $\chi^2$ function would be of the same order of magnitude as the ones when fitting the model to data sets~\textbf{I} and \textbf{II}, since data set~\textbf{V} consists of quarkonia, which are well suited for this model. Nevertheless, comparing the pull values obtained for data set~\textbf{V}, we note that almost all charmonium mesons yield positive pull values and almost all bottomonium mesons yield negative pull values, implying a skewed adjustment of the model to this data set. The smallest absolute value of the pulls from this data set is $3.1 \times 10^{-3}$ for $\chi_{b0}(1P)$, whereas the largest absolute value of the pulls is $1.6 \times 10^{-1}$ for $\chi_{c0}(1P)$. In general, the deviation in pull values is difficult to explain. It could originate from the fitting procedure being not suitable to assign the same parameter values for both charmonium and bottomonium mesons or simply by the inclusion of more data points contributing to the total value of the $\chi^2$ function. Overall, data set~\textbf{IV} fits the model the best and data set~\textbf{V} the worst.

In Table~\ref{tab:ParametersFIt}, the best-fit values for $m$, $\alpha_S$, $b$, and $\sigma$ from the fits of data sets~\textbf{I}--\textbf{V} to the model are listed. Apart from the values for $m$, which are only fitted for data sets~\textbf{I} and \textbf{II}, we observe that the values are basically divided into two groups. The first group consists of the values of data sets~\textbf{I}, \textbf{II}, and \textbf{V}, and the second one is given by the values of data sets~\textbf{III} and \textbf{IV}. The values of $\alpha_S$ for the first group are smaller (nearly by a factor of 2) than for the second group, whereas the values of $b$ and $\sigma$ for the first group are generally larger (of the order of a factor of 2) than for the second group. The reason is that the underlying mesons are different for the two groups. In the first group, the fitted mesons are quarkonia, whereas in the second group, they consist of $D$ and $B$ mesons. It is especially interesting to note that the string tension $b$ is about a factor of 2 larger for data sets~\textbf{I}, \textbf{II}, and \textbf{V} than for data sets~\textbf{III} and \textbf{IV}, thus taking into account different meson structure. Note that $\alpha_S$ and $\sigma$ obtained for data set~\textbf{II}--\textbf{V} sometimes assume the upper-end values of the intervals in which they are allowed to vary.

In Table~\ref{tab:DiquarkResults}, the predicted values for the masses of the diquarks are given, and in Table~\ref{tab:DiquarkLit}, a comparison with other works is presented. In this work, the masses of the diquarks are dependent on the parameters of the effective potential obtained from fitting data sets~\textbf{I}--\textbf{IV} to the model. Compared with the values for the diquark masses of the different works presented in Table~\ref{tab:DiquarkLit}, they deviate with at most about 250~MeV and are generally in excellent agreement with the results in Refs.~\cite{Maiani:2004vq,Debastiani:2017msn, Berezhnoy:2011xn}, which values are also predicted in the framework of nonrelativistic quark models. In Ref.~\cite{Bedolla:2019noq}, the diquark masses are studied by means of the so-called Schwinger--Dyson and Bethe--Salpeter equations, which take into account the kinetic energy as well as splittings in the spin-spin, spin-orbit, and tensor interactions. The predicted values for the diquark masses in this work are consistently smaller by about 100~MeV compared to the values in Ref.~\cite{Bedolla:2019noq}. Relativistic models, such as the ones presented in Refs.~\cite{Ebert:2005nc, Lu:2016zhe, Karliner:2016zzc, Anwar:2017toa, Anwar:2018sol}, all predict larger diquark masses, whereas models based on QCD sum rules, such as the ones in Refs.~\cite{Wang:2010sh, Kleiv:2013dta, Esau:2019hqw}, all predict smaller diquark masses (except for the $cc$ diquark mass for which Ref.~\cite{Esau:2019hqw} predicts a larger mass). The differences could be a consequence of the introduction of more and updated data in this work or relativistic effects may play a significant role, since such are not taken into account in this work.

In Table~\ref{tab:TetraquarkResults}, the resulting mass spectrum for the ground states of the tetraquarks are presented. An overall feature is that lighter tetraquarks have a larger spread in the energy eigenvalues than heavier ones, giving a larger relative difference in the masses among the states for lighter tetraquarks compared with heavier ones. In Table~\ref{tab:TetraquarkResultsComparison}, our predicted values for tetraquark masses and the corresponding ones from other works are shown. Regarding $qc\overline{qc}$ tetraquarks, the results obtained in Refs.~\cite{Patel:2014vua, Ebert:2005nc, Anwar:2018sol, Hadizadeh:2015cvx} all predict smaller masses for all states. In our model, the masses of $qc\overline{qc}$ tetraquarks are sensitive to the parameters used in the effective potential, which means that a possible explanation for this deviation could be the skew fit of the model to data set~\textbf{V}. Also, a nonrelativistic framework may not be suitable when considering heavy-light tetraquark systems, since relativistic effects play a significant role in such systems. In general, the predicted masses for $cc\overline{cc}$ tetraquarks are in good agreement with the values in Refs.~\cite{Bedolla:2019zwg, Debastiani:2017msn, Liu:2019zuc, Berezhnoy:2011xn, Barnea:2006sd, Chen:2020lgj, Wang:2019rdo, Wang:2018poa}. However, the $1^1S$ state differs by about 1~GeV in comparison to Ref.~\cite{Wu:2016vtq}. Furthermore, the predicted masses for $qb\overline{qb}$ tetraquarks are in excellent agreement with the values in Refs.~\cite{Bedolla:2019zwg, Ebert:2005nc, Hadizadeh:2015cvx}, and the relative deviation among the masses of different tetraquark states is overall small for this type of tetraquarks. Concerning $bb\overline{bb}$ tetraquarks, the predicted masses are in very good agreement with Refs.~\cite{Bedolla:2019zwg, Anwar:2017toa, Berezhnoy:2011xn, Wang:2018poa}, but consistently smaller by about 0.5~GeV--1.0~GeV compared to the values in Refs.~\cite{Liu:2019zuc, Wu:2016vtq, Wang:2019rdo}. For the heavy tetraquarks (i.e., $cc\overline{cc}$ and $bb\overline{bb}$), the predicted masses obtained in Refs.~\cite{Liu:2019zuc, Wu:2016vtq, Wang:2019rdo} are consistently and significantly larger than those obtained in this work.  In Ref.~\cite{Liu:2019zuc}, the color-magnetic interaction is adopted to calculate the masses, and in Refs.~\cite{Wu:2016vtq, Wang:2019rdo}, a similar model to the one used in this work is considered, but the variational principle is applied when solving the Schr{\"o}dinger equation. This difference in the modeling approach could be the reason for the differences in the results.

Finally, we compare our results to results from hadronic molecular systems based on lattice QCD, which results are summarized in, e.g., Refs.~\cite{Esposito:2016noz,Guo:2017jvc}. In Ref.~\cite{Tornqvist:2004qy}, masses for $D^* \bar{D}^*$ and $B^* \bar{B}^*$ have been computed in a one-pion exchange model to be about 4015~MeV and 10600~MeV, respectively, which, in fact, are in good agreement with our masses for $1S$~$qc\overline{qc}$ (4080~MeV--4260~MeV) and $1S$~$qb\overline{qb}$ (10400~MeV--10500~MeV) tetraquarks; see Table~\ref{tab:TetraquarkResultsComparison}. Furthermore, in Ref.~\cite{HidalgoDuque:2012pq}, masses for $[q\bar{c}][\bar{q}c]$ molecular states with $1^{+-}$ and $2^{++}$ are computed in the interval 3820~MeV--4010~MeV, which could be compared to our results for masses of $qc\overline{qc}$ tetraquarks in the interval 4160~MeV--4260~MeV that is slightly higher. In Refs.~\cite{Chiu:2006hd,Padmanath:2015era}, evidence for $X(3872)$ with $1^{++}$ has been found within about 50~MeV for its mass, which is much closer to the experimental mass than our mass for the corresponding $qc\overline{qc}$ tetraquark state (although with $1^{+-}$), which is much larger. Then, in Ref.~\cite{Voloshin:2013dpa}, the mass for $Z_c(3900)$ as $D \bar{D}^*$ has been estimated to approximately 4030~MeV, which is closer to the experimental mass than our mass for the $qc\overline{qc}$ tetraquark state with $1^{+-}$, which is about 4160~MeV. Next, in Ref.~\cite{Chen:2016lkl}, the mass for $Z(4430)$ has been found to be about 4475~MeV, which should again be compared to our mass for the $qc\overline{qc}$ tetraquark state with $1^{+-}$. In Ref.~\cite{Chen:2015ata}, masses for $[q\bar{c}][\bar{q}c]$ and $[q\bar{b}][\bar{q}b]$ molecular states with $1^{+-}$ are computed in the intervals 3850~MeV--4220~MeV and 9920~MeV--10500~MeV, respectively, which can be compared to our results of 4160~MeV and 10500~MeV, respectively, that lie in those intervals. For masses of $qb\overline{qb}$ tetraquarks, a mass difference between $Z_b(10610)$ and $Z_b(10650)$ of about 46~MeV has been obtained \cite{Bondar:2011ev}, which should be compared to our zero result for this mass difference (see the discussion in the next subsection). In addition, a mass of about 10600~MeV for $Z_b(10610)$ as $B^* \bar{B}^*$ with $1^{+-}$ has been obtained \cite{Guo:2013sya}, which is similar to our corresponding result of about 10500~MeV.

\subsection{Comparison with experimental results}

Considering experimental results, there are about ten candidates of tetraquarks that are listed in the particle listings of the Particle Data Group \cite{Zyla:2020}. These experimental tetraquark candidates are $\chi_{c1}(3872)$ [$1^{++}$], $Z_c(3900)$ [$1^{+-}$], $X(3915)$ [$0^{++},2^{++}$], $X(4020)^\pm$ [$?^{?-}$], $\chi_{c1}(4140)$ [$1^{++}$], $\chi_{c1}(4274)$ [$1^{++}$], $\psi(4360)$ [$1^{--}$], $Z_c(4430)$ [$1^{+-}$], and $\psi(4660)$ [$1^{--}$], which are all potential $qc\overline{qc}$ tetraquarks, and $Z_b(10610)$ [$1^{+-}$] and $Z_b(10650)$ [$1^{+-}$], which are both potential $qb\overline{qb}$ tetraquarks. There exists a classification of tetraquark states based on ``good'' diquarks ($N^{2S+1} L_J = 1^3 S_1$) such that \cite{Ebert:2008se, Anwar:2018sol, Bedolla:2019zwg}
\allowdisplaybreaks
\begin{align}
J^{PC} = 0^{++} \quad &\to \quad {}^1S_0, {}^5D_0 \,, \nonumber\\
J^{PC} = 0^{-+} \quad &\to \quad {}^3P_0 \,, \nonumber\\
J^{PC} = 0^{--} \quad &\to \quad - \,, \nonumber\\
J^{PC} = 1^{++} \quad &\to \quad {}^5D_1 \,, \nonumber\\
J^{PC} = 1^{+-} \quad &\to \quad {}^3S_1, {}^3D_1 \,, \nonumber\\
J^{PC} = 1^{-+} \quad &\to \quad {}^3P_1 \,, \nonumber\\
J^{PC} = 1^{--} \quad &\to \quad {}^1P_1, {}^5P_1 \,, \nonumber\\
J^{PC} = 2^{++} \quad &\to \quad {}^5S_2, {}^1D_2, {}^5D_2 \,, \nonumber\\
J^{PC} = 2^{+-} \quad &\to \quad {}^3D_2 \,, \nonumber\\
J^{PC} = 2^{-+} \quad &\to \quad {}^3P_2 \,, \nonumber\\
J^{PC} = 2^{--} \quad &\to \quad {}^5P_2 \,. \nonumber
\end{align}
Summarizing the current experimental situation, we observe that the following five states are present $0^{++}$, $1^{++}$, $1^{+-}$, $1^{--}$, and $2^{++}$, which means that we have the following ten spectroscopic states of tetraquarks: ${}^1S_0$, ${}^5D_0$, ${}^5D_1$, ${}^3S_1$, ${}^3D_1$, ${}^1P_1$, ${}^5P_1$, ${}^5S_2$, ${}^1D_2$, and ${}^5D_2$ to investigate. Furthermore, if one considers spin-1 diquarks and antidiquarks (i.e., axial-vector diquarks), then one has only three possibilities for the total spin $S = 0,1,2$ of the tetraquarks, i.e., three wave functions for each state $N^{2S+1}L$: $N^1S$, $N^3S$, $N^5S$, $N^1P$, $N^3P$, $N^5P$, $N^1D$, $N^3D$, and $N^5D$, which are nine possibilities \cite{Debastiani:2017msn}. Note that the total angular momentum quantum number $J$ is dropped from the spectroscopic states of the tetraquarks, since our model is independent of $J$. Therefore, we should compute the following eight interesting states: $1^1S$, $1^3S$, $1^5S$, $1^1P$, $1^5P$, $1^1D$, $1^3D$, and $1^5D$ (i.e., not $1^3P$ included), which are all ground states ($N = 1$), and compare the experimental values for the masses of tetraquarks with our theoretically predicted values of the masses using the allowed ground states for each tetraquark candidate. In comparing our theoretical predications in Table~\ref{tab:TetraquarkResults} with the experimental values for the tetraquark masses (cf.~Ref.~\cite{Zyla:2020}), we find the following agreement within 100~MeV for $qc\overline{qc}$ tetraquarks:
\begin{align}
1^1P: 4582~{\rm MeV} \quad &\leftrightarrow \quad \psi(4660) \; [(4633 \pm 7)~{\rm MeV}] \,, \nonumber\\
1^5P: 4591~{\rm MeV} \quad &\leftrightarrow \quad \psi(4660) \; [(4633 \pm 7)~{\rm MeV}] \,, \nonumber
\end{align}
and within about 250~MeV for $qc\overline{qc}$ tetraquarks,
\begin{align}
1^3S: 4156~{\rm MeV} \quad &\leftrightarrow \quad Z_c(3900) \,, \nonumber\\
1^1S: 4076~{\rm MeV} \quad &\leftrightarrow \quad X(3915) \,, \nonumber\\
1^1P: 4582~{\rm MeV} \quad &\leftrightarrow \quad \psi(4360) \,, \nonumber\\
1^5P: 4591~{\rm MeV} \quad &\leftrightarrow \quad \psi(4360) \,, \nonumber\\
2^3S: 4693~{\rm MeV} \quad &\leftrightarrow \quad \psi(4360) \,, \nonumber
\end{align}
where $2^3S$ is an excited state, and for $qb\overline{qb}$ tetraquarks,
\begin{align}
1^3S: 10472~{\rm MeV} \quad &\leftrightarrow \quad Z_b(10610) \,, \nonumber\\
1^3S: 10472~{\rm MeV} \quad &\leftrightarrow \quad Z_b(10650) \,. \nonumber
\end{align}
Thus, it seems that the most likely tetraquark candidate to be described with our model is $\psi(4660)$ as either a $1^1P$ state of mass 4582~MeV or a $1^5P$ state of mass 4591~MeV (see Table~\ref{tab:TetraquarkResults}). Furthermore, the $qc\overline{qc}$ tetraquark candidates $Z_c(3900)$, $X(3915)$, and $\psi(4360)$ as well as the $qb\overline{qb}$ tetraquark candidates $Z_b(10610)$ and $Z_b(10650)$ could be described by our model. Finally, the $cc\overline{cc}$ and $bb\overline{bb}$ tetraquarks are interesting objects to study in the sector of exotic hadrons. The results obtained in this work suggest that the mass of the fully charmed tetraquark could be about 5960~MeV or above in its ground state, whereas the mass of the fully bottom tetraquark could be as large as 18720~MeV (see Tables~\ref{tab:TetraquarkResults} and \ref{tab:TetraquarkResultsComparison}).

\section{Summary and conclusions}
\label{sec:summary}

We have investigated a model of tetraquarks, assumed to be compact and to consist of diquark-antidiquark pairs, in a nonrelativistic framework and predicted mass spectra for the $qc\overline{qc}$, $cc\overline{cc}$, $qb\overline{qb}$, and $bb\overline{bb}$ tetraquarks. Considering tetraquarks as bound states of axial-vector diquarks and antidiquarks, a simple model originally formulated for quarkonia has been adopted and used to calculate and predict the masses of different tetraquark states. For the first time, a total number of 45 mesons, including both charm and bottom quarks, and the most recent corresponding data on the masses of these mesons \cite{Zyla:2020} have been used to fit the free parameters of the model. Particularly, we have found predictions for four axial-vector diquark masses, and subsequently, a total number of 24 tetraquark masses, which are all presented in Tables~\ref{tab:DiquarkResults} and \ref{tab:TetraquarkResults}. In comparison with other nonrelativistic models, our results for the $cc\overline{cc}$, $qb\overline{qb}$, and $bb\overline{bb}$ tetraquarks are shown to be in excellent agreement with earlier results presented in the literature. However, considering $qc\overline{qc}$ tetraquarks, our results deviate slightly from earlier results and the predicted masses of these tetraquarks are consistently larger than the ones found in the literature. For the masses of heavy-light tetraquark states, i.e., $qc\overline{qc}$ and $qb\overline{qb}$, we have been able to identify some of these states with experimentally proposed tetraquark candidates. One such identification includes the $\psi(4660)$ tetraquark candidate, which can be proposed to be the $qc\overline{qc}$ tetraquark in either the state $1^1P$ or the state $1^5P$. For $qb\overline{qb}$ tetraquarks, the tetraquark candidates $Z_b(10610)$ and $Z_b(10650)$ could both be identified with the state $1^3S$. Concerning the heavy tetraquark states, i.e., $cc\overline{cc}$ and $bb\overline{bb}$, the model predicts the mass of the fully charmed tetraquark to be 5960~MeV and the mass of the fully bottom tetraquark to be 18720~MeV, both values correspond to values obtained for their respective ground states. Finally, our model could also be used to predict masses for other potential tetraquark states for which no experimental data exist today.

\begin{acknowledgments}
We would like to thank Franz F.~Sch{\"o}berl for providing us with his \emph{Mathematica} notebook that inspired the code used in this work for solving the Schr{\"o}dinger equation.
T.O.~acknowledges support by the Swedish Research Council (Vetenskapsr{\aa}det) through Contract No.~2017-03934.
\end{acknowledgments}

\providecommand{\noopsort}[1]{}\providecommand{\singleletter}[1]{#1}%

\end{document}